\documentstyle[aps,pra,twocolumn,epsfig]{revtex}
\begin{document}
\title{\large \bf Two-particle correlations via quasi-deterministic analyzer model}
\author{Bill Dalton\thanks{Email: bdalton@stcloudstate.edu}}
\address{Department of Physics, Astronomy and Engineering Science\\
St. Cloud State University\\
St. Cloud, MN 56301, USA}
\date{\today}
\maketitle
\begin{abstract}
We introduce a quasi-deterministic eigenstate transition model of
analyzers in which the final eigenstate is selected by initial
conditions. We combine this analyzer model with causal spin
coupling to calculate both proton-proton and photon-photon
correlations, one particle pair at a time. The calculated
correlations exceed the Bell limits and show excellent agreement
with the measured correlations of [M. Lamehi-Rachti and W. Mittig,
Phys. Rev. D 14 (10), 2543 (1976)] and [ A. Aspect, P.  Grangier
and G. Rogers, Phys. Rev. Lett. 49 91 (1982)] respectively. We
discuss why this model exceeds the Bell type limits.
\end{abstract}

% body of paper here

 Past efforts to explain two-particle correlation measurements with hidden variable
theories combined with probability methods have failed
\cite{Af-Sell,Belinf,Ballen}. Could we successfully explain
measured correlations by replacing the probability feature by a
deterministic decision process, accumulating the distributions one
particle pair at a time? A deterministic trajectory
model\cite{Dal-G} has been previously used to explain "ghost
diffraction" patterns \cite{Shih-G}. In a separate more extensive
paper \cite{Dal-Malus}, we use a quasi-deterministic analyzer
model, to explain the law of Malus for in-sequence photon
counting. Here, we use this model in detailed calculations to
explain the proton-proton correlations of \cite{LR-M}, as well as
the four-angle photon-photon correlations of Aspect et al.
\cite{Aspect-81,Aspect-82}.

Polarized beam splitters as well as Stern-Gerlach type analyzers
for spinors are eigenvalue splitters. The quasi-deterministic
model outlined here is based on eigenvalue selection.  To describe
this theory we first rewrite certain relations contained in the
generic two-component eigenvalue equation in terms of a convenient
Stokes variable representation. Stokes representations for both
spinors and vectors are similar and well developed \cite{McMas}.

Consider a Hermitian matrix with elements $h_{11} = a$,
$h_{22}=d$, $h_{12}=h\exp(-\imath\phi)$, and
$h_{21}=h\exp(\imath\phi)$, where $a$, $d$, $h$, and $\phi$ are
real. With $\Psi^T= {\left[\Psi_x,\Psi_y\right]}$,
$\Psi_x=A_x\cos(\beta)\exp(\imath\alpha_x)$ and
$\Psi_y=A_y\sin(\beta)\exp(\imath(\alpha_x+\delta)$, we can
extract from the eigenvalue equation $h\Psi  =  \lambda\Psi$ the
following relations \cite{Dal-Malus}.
\begin{equation}\label{eig1}
\textbf{\textsl{P}}\cdot\textbf{\textsl{S}}=\pm P_0S_0
\end{equation}
\begin{equation}\label{eig2}
S_3P_2-S_2P_3=0
\end{equation}
\begin{equation}\label{eig3}
S_1P_1=\pm(P_1)^2S_0/P_0
\end{equation}
Here, $S$ and $P$ are Stokes sets that represent the field and
matrix respectively and the two signs are associated with the two
eigenvalues. The components of $S$ have the standard form $S_0 =
\Psi^\dagger\Psi$, $S_1 = S_0\cos(2\beta) $, $S_2 =
S_0\sin(2\beta)\cos(\delta)$, and $S_3
=S_0\sin(2\beta)\sin(\delta)$. The components of $P$ are given by
$P_0 = \sqrt{(a-d)^2+4h^2}$, $P_1 = P_0\cos(2\alpha)$, $P_2 =
P_0\sin(2\alpha)\cos(\phi)$ and $P_3 =P_0\sin(2\alpha)\sin(\phi)$,
where $\tan(2\alpha)=2h/(a-d)$ and the eigenvalues are given by
$\lambda_\pm = (a+d \pm P_0)/2$. Equation (\ref{eig2}) means
$\sin(\delta-\phi)=0$.

The vectors $\textbf{\textsl{S}}$ and  $\textbf{\textsl{P}}$
represent points on the Poincare' polarization sphere. If an
incident field is to make a transition to an eigenstate, equations
(\ref{eig1}), (\ref{eig2}), and (\ref{eig3}) must be satisfied.
From (\ref{eig3}) we see that the eigenstate condition is related
to the sign of the product $S_1P_1$. The Stokes variables rotate
with twice the rotation angle of the field components.

The traditional \textit{e} and \textit{o} rays of classical
macroscopic electro-optics correspond to two points on opposite
sides of the Poincare' sphere indicated by $P_1=1$ and $S_1=\pm1$
for one crystal type. The two eigenvalues, on which the separation
decision is made, determine the $P$ sphere radius via $\lambda_+
-\lambda_- = P_0$, but not the direction of $\textbf{\textsl{P}}$.
For a fixed relative phase $\phi$, this gives one free variable
for the matrix Stokes vector.

The point of view here is that the classical matrix Stokes vector
used for a macroscopic field of many photons does not necessarily
represent the matrix Stokes vector experienced by individual
photons. The first assumption of this model (The distributive
assumption) is that the $\textbf{\textsl{P}}$ vectors experienced
by individual incident pulses are distributed in the one free
variable, at least for the surface transition region. For a given
incident pulse, we randomly select this degree of freedom of
$\textbf{\textsl{P}}$ from a distribution described below.

The second assumption of this model (The deterministic assumption)
is that the incident pulse makes a transition to one eigen-channel
or the other, and that the choice is made with a deterministic
criteria based on initial conditions of the incident pulse at the
analyzer and the randomly selected $\textbf{\textsl{P}}$. From
(\ref{eig2}) we see that the sign of $S_1P_1$ indicates the
eigen-channel choice. We indicate this product as a function of
$z$, the depth into the crystal, as follows.
\begin{equation}\label{T0l}
T(z)=S_1(z)P_1(z)
\end{equation}
The deterministic criteria on which the calculations here are
based is that the sign of $T(z)$ after the transition is the same
as, and determined by the sign of $T(0)$. For a given value for
$S_1(0)$ and $P_1(0)$ we make the eigen-channel decision by
testing the sign of $T(0)$. Are the residual deviations of
$P_1(z)$ from the macroscopic field value only surface affects,
rapidly decreasing with depth into the crystal, or, do they extend
throughout the crystal? Both cases give the same correlations, and
as shown in \cite{Dal-Malus}, both cases give rise to the law of
Malus. To correctly describe the correlations, or the law of
Malus, it is only necessary that the surface value $P_1(0)$ is
selected from a distribution. Because of averaging, it would be
unreasonable to expect to answer this distribution question with
experiments using a macroscopic field of many photons.

For a frame attached to the crystal analyzer (say aligned with the
$\textit{e}$ and $\textit{o}$ rays), we can partition the matrix
Poincare' sphere with hemispheres indicated by the sign of $P_1$.
However, viewed from a frame attached to one analyzer, this
partition for the other analyzer is rotated. With respect to an
analyzers attached frame, we choose $\textbf{\textsl{P}}$ by
choosing $2\alpha$ via $2\alpha=arg$ where $arg =
\arccos(u)-\pi/2$ and $u$ is selected uniformly from the interval
$[-1,1]$. For our linear polarizer, we have $\cos(\phi)=\pm1$ in
the crystal frame. Viewed from a frame fixed to the first
analyzer, the $\textbf{\textsl{P}}'$ hemisphere axis for the
second analyzer is rotated from the first by $q=\theta$ for
spinors and $q=2\theta$ for vectors. In this fixed reference
frame, sampling for the second analyzer is made using
$2\alpha'=arg(u') \pm q$ where $u'$ is selected uniformly from the
interval $[-1,1]$, but independent of $u$.  The independence of
the selection of $u$ and $u'$ represents a local stochastic
element of this theory, and is the reason why we call this a
quasi-deterministic model.

Pair counts for the four different coincidence combinations are
represented here by $N_{++}$ ,$N_{+-}$ , $N_{-+}$ and $N_{--}$ .
For instance $N_{+-}$ is the count for the number of pairs with a
$(+)$ for the first analyzer an $(-)$ for the second analyzer. The
four pair-counts are accumulated, one pair at a time. The
correlations are calculated using the function
$\gamma(\theta)=(N_{++}+ N_{--}-N_{+-}-N_{-+})/N$ \cite{Garucc},
where $ N =N_{++}+ N_{--}+N_{+-}+N_{-+}$ is the total number of
pairs counted.

    The proton-proton correlations in \cite{LR-M} were measured using two
protons in coincidence produced via scattering a proton beam by a
hydrogen target. Only a small percent (We assume $3\%$ for the
calculations reported here.) of these proton pairs are in the
triplet state. Information about the spin correlations of the
protons was obtained in \cite{LR-M} via scattering from a carbon
target rather than using a Stern-Gerlach apparatus in each beam.
The two protons each are incident upon carbon targets. Four
detectors are used, two after each carbon target. On one side,
the two detectors are positioned in the reaction plane. On the
other, the two detectors are rotated out of the reaction plane by
an angle $\theta$. This is an azmuthal rotation around the
incident beam direction on that side.

 For our reference side, we choose a $(1-2-3)$ right
handed coordinate system with $P_3$  in the beam direction, and
$P_2$ perpendicular to the plane of the two detectors. With this,
one detector is in the $P_1\geq0$ hemisphere and the second
detector is in the other. With a similar choice for the other
analyzer, its $P_1' \geq0$ hemisphere seen in the fixed reference
frame is rotated by $\theta$ around the beam axis. This comparison
is made after a rotation to account for the non-alignment of the
proton beams. For convenience, we use unit normalization and
$S_1(0)=\cos(2\beta)$ for the spinor incident on the first
analyzer side, where $2\beta$ is sampled uniformly from the
interval $[0,2\pi)$ . For the triplet case, we set $S_{1}'(0)=
S_1(0)$, and for the singlet case, we set $S_{1}'(0)= -S_1(0)$.
This represents the causal coupling between the two sides.

Calculations are made using about $10000$ pairs at each relative
angular setting. For each member of a pair of proton spinor
states, the deterministic decision at the analyzer is made via
testing the sign of $T(0)$ at each analyzer. For this, recall that
$T'(0)$ , given in (\ref{T0l}), has two factors each of which can
have either sign for the rotated analyzer.

\begin{figure}[htbp]
  \vspace*{8.00cm}
  \hspace*{6.00cm}
  \includegraphics{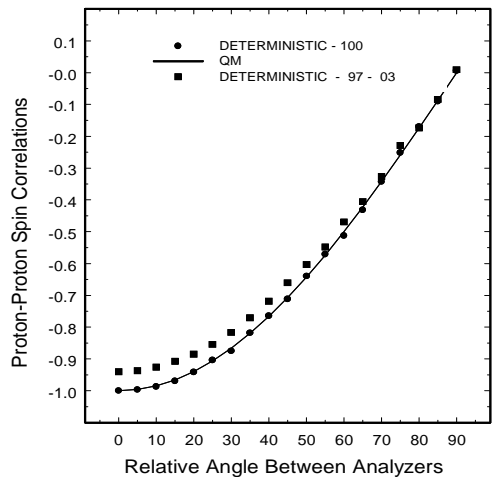}
% \centerline{\epsfxsize 6in \epsfbox{eprspin.eps}}
 \vskip -0.5in
    \caption{Proton-Proton Spin Correlations: Solid circles (Solid squares)
    indicated deterministic correlations calculated assuming $100\%$ ( $97\%$ )
    orthogonal states; Solid line indicates the $-\cos(\theta)$
    formula of quantum mechanics.}
    \label{f1}
\end{figure}

Results of calculations assuming that $100\%$ of the proton pairs
are mutually orthogonal are shown by the solid circles in
Fig.~\ref{f1}. The solid line corresponds to the formula
$-\cos(\theta)$ of quantum mechanics for the same case. The
deterministic correlations calculated with $3\%$ of the incident
pairs parallel and $97\%$ orthogonal are shown by the solid
squares in Fig.~\ref{f1}. This curve falls well within the error
bars of the measurements of \cite{LR-M}.  The proton-proton
correlations calculated with this deterministic analyzer model
agree well with results of quantum mechanics and with the
experimental data \cite{LR-M}.

Here we present calculations for four-angle photon-photon
correlation \cite{Aspect-81,Aspect-82} in which the two analyzers
are set at angles $(a,a')$ and $(b,b')$. These angles are chosen
to give a relative angle of $\theta$ for two of the combinations,
$-\theta$ for another combination, and $3\theta$ for the fourth.
Each of the four angular settings represents an independent
measurement. Here, the correlation function $\gamma(\theta)$ for
each case is calculated as described above and the four-angle
correlation function is then calculated by
$\gamma_4(\theta)=2\gamma(\theta)+\gamma(-\theta)-\gamma(3\theta)$.
For the input light, we use an isotropic distribution around the
beam axis, sampling $2\beta$ uniformly from $[0,2\pi]$. We use
$S_{1}'(0)= S_1(0)$ for the causal coupling.

\begin{figure}[htbp]
  \vspace*{8.5cm}
  \hspace*{6.00cm}
  \includegraphics{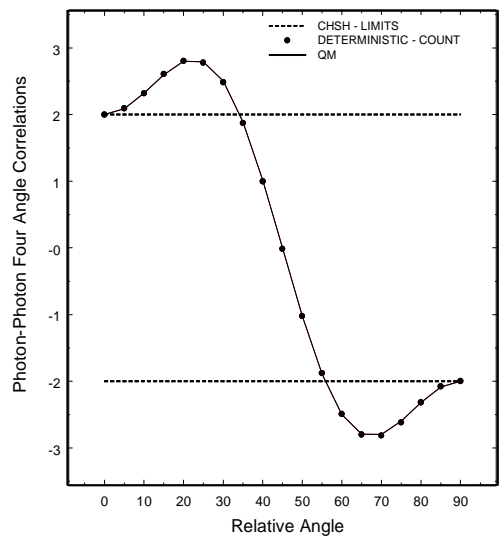}
%  \centerline{\epsfxsize 6in \epsfbox{eprlight4.eps}}
  \vskip -.5in
    \caption{Four-Angle Photon-Photon Polarization Correlations: Solid
    circles indicated deterministic correlations calculated one photon
    pair at a time; Solid line indicates the $3\cos(2\theta)-\cos(6\theta)$
    formula of quantum mechanics.}
    \label{f2}
\end{figure}

 In  Fig.~\ref{f2} the solid circles indicate
the calculated four-angle correlation, and the solid line
represents the function $3\cos(2\theta)-\cos(6\theta)$ of quantum
mechanics. The two dash lines represent the well known (CHSH)
limits \cite{CHSH}. We emphasize that all contributions to each
$\gamma(\theta)$ were calculated using the four pair counts only
so there is no chance that an accidental scale factor can be
included. The calculated results presented in Fig.~\ref{f2}
clearly confirm that the four-angle photon-photon correlations can
be explained using this causal quasi-deterministic analyzer model.

In contrasting the above results with previous studies and views
on this topic, many questions are raised. In discussions with many
colleagues, the questions seem to focus into three. \textsl{First,
if we replaced the deterministic decision at each analyzer with
one based on probability, would we get the same results? }The
answer is no! In studies leading to this work, the author has
tried many different hidden variable probability models, and not
found one, no matter what sampling distribution used, that agrees
with the deterministic model and quantum mechanics. This finding
seems to be consistent with the many studies of probability models
by previous authors \cite{Belinf,Greenb}.

We have the second question. \textsl{Since this theory is based on
individual photons making transitions to one or the other
eigenvalue channel, is it consistent with the Law of Malus? }To
the author's mind, this question is more important than asking if
this theory can explain photon-photon correlations. We recall that
for distributions accumulated from individual photon counting, the
law of Malus is postulated as a fundamental starting point. The
analysis of this question is the central theme of a separate and
more extensive study by the author \cite{Dal-Malus} . The answer
is that this theory is not only consistent with the law of Malus,
but as shown in  \cite{Dal-Malus}, it gives rise to it for
in-sequence photon counting. We must take double serious any
theory that explains both the observed photon-photon correlations
and the law of Malus.

A third question is asked. \textsl{Why do these calculations not
respect the Bell limits, in particular the CHSH limits \cite{CHSH}
for the four-angle correlation ?} First, given the assumptions
used, these limits follow rigorously \cite{CHSH}. The answer is
that this deterministic model does not respect some assumptions on
which this Bell type theorem is based. To elucidate this, consider
the analyzer functions
$T(0)=S_1(0)P_1(0)=\cos(2\beta)\cos(arg(u))$ and
$T'(0)=S_1(0)P_1'(0)=\cos(2\beta)\cos(arg(u') \pm 2\theta)$. The
independent parameters $u$ and $u'$ represents local stochastic
features of this theory. Within the distribution range, and
depending on the angle $2\theta$ , there are certain values of
$u'$ for which the second factor in $T'(0)$ can be positive, and
certain values for which this factor is negative. Because of the
stochastic variable $u$ , we cannot know the path of the
individual quantum pulse until after we have measured it. However,
we still have distribution information. The number of times the
second factor in $T'(0)$ is positive depends on the angle
$2\theta$. The causal link for individual photon pairs is
contained in the first two factors of $T(0)$ and $T'(0)$. As  a
consequence, we have correlations in the distributions which
depend on the relative angle. The presence of the local stochastic
variable, a necessary part of this theory, does not destroy all
causal correlation information in the distributions.

 Using
$\gamma=\sum{A_iB_i}/N$, with ${A_i=\pm1}$ and ${B_i=\pm1}$, where
$A_i$ and $B_i$ represent the signs of $T(0)$ and $T'(0)$
respectively, we can easily arrive at the following inequality.
 \begin{eqnarray}
 \left|\gamma_4\right|  \leq  \sum{\left|B_i(a-b)+B'_i(a-b')\right|}/N\nonumber \\
 +\sum{\left|B_i(a'-b)-B'_i(a'-b')\right|}/N
\end{eqnarray}
Because of the angular dependence of the second factor in $T'(0)$,
the selection of $B_i$ depends on the relative angle. If we did
not have this relative angular dependence, this inequality would
reduce to $\left|\gamma_4\right|\leq2$ of \cite{CHSH}.  This angle
dependence does not mean that what happens at one analyzer
influences what happens at the other. Rather, the angle dependence
is created by the causal coupling relative to the analyzer
settings.

In this model calculations, we make no use of entangled states.
Our calculated results depend on the causal coupling, the
deterministic decision at the analyzer and the stochastic matrix
model. If we leave one of these out, such as replacing the
deterministic process at the analyzer with probability, we can not
describe the data. This is a simple theory which can explain the
law of Malus for in-sequence photon counting as well as explain
some well known two-particle correlations.

\vskip 1.0in
%\begin{figure}{htbp}
% \epsfbox{samp100.eps}
%  \caption{My baby}
%  \label{}
%\end{figure}
\vskip .5in
% \begin{figure}[h]
% \vspace*{6.00cm}
%\hspace*{6.00cm}
% \special{psfile=samp100.eps
%        vscale=80 hscale=80 angle=0 hoffset=-20 voffset=20}
%\end{figure}

%\begin{figure}[htbp]
%  \begin{center}
% \epsfysize 120mm
% \epsfxsize 90mm
 % \epsfbox{eprspin.eps}
%  %  \caption{Proton-Proton Spin Correlations: Solid circles (Solid squares)
 %   indicated deterministic correlations calculated assuming $100\%$ ( $97\%$ )
    %orthogonal states; Solid line indicates the $-\cos(\theta)$
   % formula of quantum mechanics.}
  % \label{f3}
 % \end{center}
%\end{figure}

%\begin{figure}[htbp]
 % \begin{center}
% \epsfxsize 70mm
 % \epsfbox{eprlight4.eps}
 %  \caption{Four-Angle Photon-Photon Polarization Correlations: Solid
 %   circles indicated deterministic correlations calculated one photon
 %   pair at a time; Solid line indicates the $3\cos(2\theta)-\cos(6\theta)$
  %  formula of quantum mechanics.}
 %  \label{f3}
 % \end{center}
%\end{figure}
\end{document}